\documentclass[aps,floatfix,preprint,nofootinbib]{revtex4}

\usepackage{graphicx}
\usepackage{epic}
\usepackage{eepic}
\usepackage{latexsym}

\newcommand{\eq}[1]{(\ref{#1})}
\newcommand{\be}{\begin{equation}}
\newcommand{\ee}{\end{equation}}
\newcommand{\bea}{\begin{eqnarray}}
\newcommand{\eea}{\end{eqnarray}}

\newcommand{\vs}[1]{\vspace{#1 mm}}
\newcommand{\hs}[1]{\hspace{#1 mm}}

\def\h{\hat}

\def\a{\alpha}

\def\cc{\gamma}

\def\d{\delta}
\def\D{\Delta}
\def\e{\epsilon}

\def\fr{\frac}

\def\h{\eta}

\def\l{\lambda}

\def\m{\mu}
\def\n{\nu}

\def\r{\rho}
\def\s{\sigma}

\def\th{\theta}
\def\Th{\Theta}

\def\O{\Omega}

\def\o{\omega}

\def\del{\partial}

\let\bm=\bibitem
\def\nn{\nonumber}

\newcommand{\dpo}{\delta p_0}
\newcommand{\dpi}{\delta p_i}

\begin{document}

\title{Null geodesic congruences, gravitational lensing and CMB intensity profile}
\author{Ali Kaya}
\affiliation{\vs{3}Department of Physics, Astronomy and Geophysics, Connecticut College, New London, CT 06320, USA \vs{5}}

\begin{abstract}
	
It is known that the Cosmic Microwave Background (CMB) temperature fluctuations are modified by gravitational lensing since the angular positions of photons are altered by the metric perturbations. We reconsider this effect in the context of geodesic deviation on a perturbed Friedmann-Robertson-Walker (FRW) spacetime with scalar and tensor modes. We first give an alternative derivation of the Sachs-Wolfe effect by using the solutions of the null geodesic equation. Then, we determine the two-dimensional induced metric on the transverse cross section of the null geodesic congruence corresponding to a beam of photons after decoupling. This metric, whose variation along the congruence can be decomposed as expansion and shear, is shown to produce a nontrivial intensity profile which can be characterized by three variables analogous to the Stokes parameters of the polarization tensor. 

\end{abstract}

\maketitle

\section{Introduction}

In general relativity the world lines of freely falling particles in a gravitational field are geodesics. This assumption, which is directly related to the equivalence principle, is crucial for the physical viability of the theory (one can show that the stress-energy-momentum conservation for a sufficiently small body having sufficiently weak self gravity also implies the geodesic motion \cite{geroch}). Studying the evolution of families of nearby geodesics gives invaluable geometrical information as well. The singularity theorems came out as a result of this study by analyzing the evolution of time-like and null geodesic congruences using Raychaudhuri's equation. It is important to note that in the field theory description, for instance when the Maxwell's equations are considered on a curved spacetime, the geodesic motion arises in the geometrical optics approximation. 

Understanding the null geodesic propagation is also essential in cosmology. Particularly, it is important to determine in detail the evolution of freely propagating CMB photons after decoupling. Since the CMB photons were released in the state of thermal equilibrium, one introduces a phase space distribution function to describe their physics and apply Liouville's theorem by interpreting the null geodesic evolution in the Hamiltonian framework. This leads to the Sachs-Wolfe effect and helps one to relate the observed CMB temperature fluctuations to the primordial power spectrum of cosmological perturbations. 

It is known that CMB polarization data contain crucial cosmological information. CMB photons were polarized during recombination as a result of Thomson scattering where the cross section depends on the incoming and outgoing photon polarizations. For an initially unpolarized distribution of electromagnetic radiation, the main polarization effect arises from the quadruple moment of the photon distribution function. The evolution of the whole system is described by the coupled Boltzmann equations. Since the polarization tensor depends on the basis vectors chosen on the sphere, one has to introduce spin weighted spherical harmonics for their expansion \cite{p1,p2,p3}. One can define the so called $E$ and $B$ modes of polarizations as spin weight zero scalar quantities where only the tensor modes give rise to the $B$ type polarization as a crucial distinctive feature (see \cite{p4} for a pedagogical review). 

\begin{figure}
\centerline{\includegraphics[width=9cm]{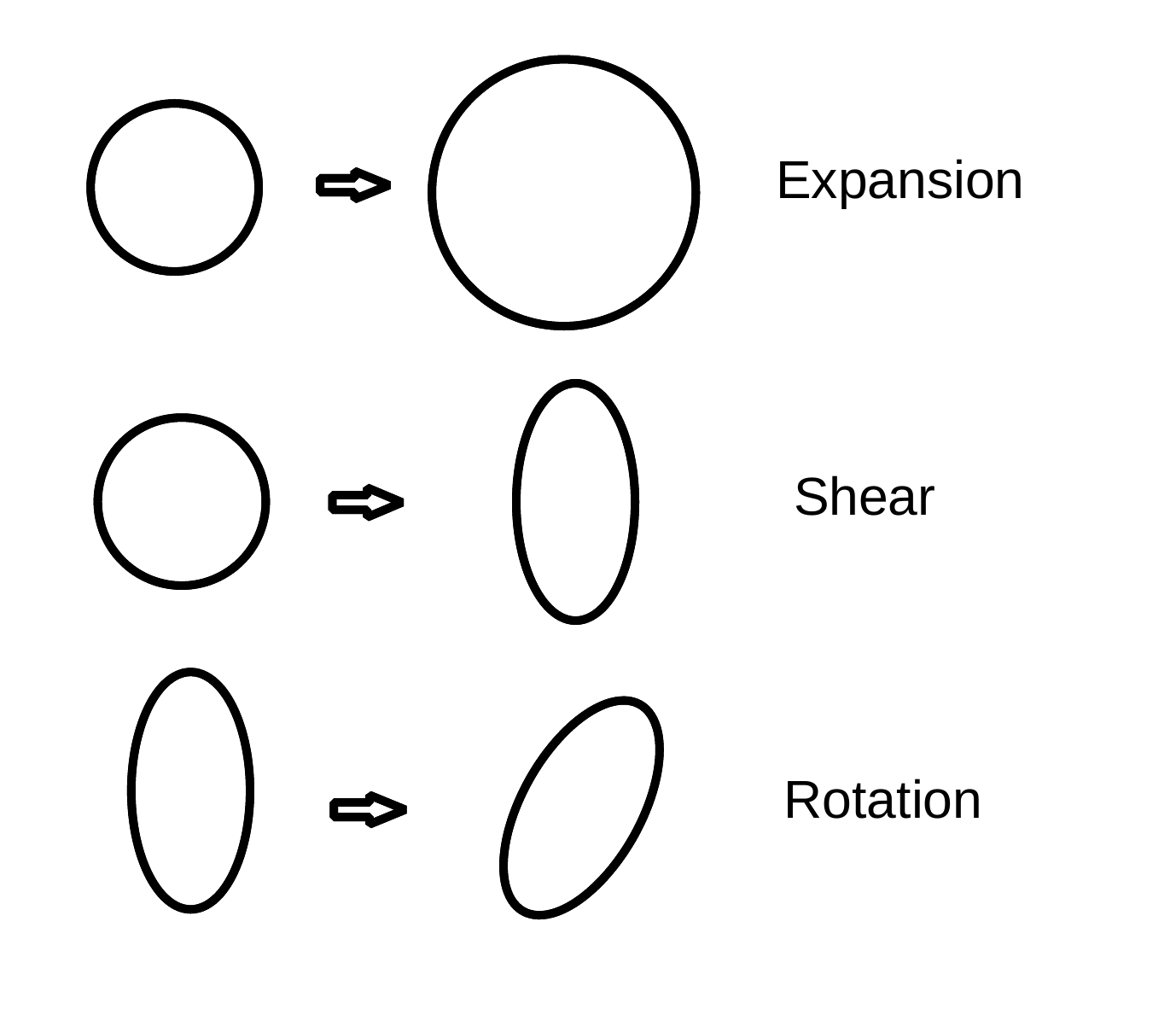}}
\caption{A schematic illustration of the expansion, shear and rotation of a geodesic congruence showing how the cross sectional area evolves.} 
\label{fig1}
\end{figure}

After decoupling, CMB photons move along null geodesic congruences and it is possible to determine the corresponding expansion, shear and rotation parameters, which characterize how a specified two-dimensional cross sectional area evolves along a family. Specifically, while the expansion equals the growth rate of the area, the shear and rotation parameters quantify how the shape is deformed and rotated, respectively, see Fig. \ref{fig1}. 
 
The intensity of a photon beam is inversely proportional to its cross sectional area so it decreases when the congruence expands. One aim of this work is to quantify this observation for the CMB temperature fluctuations after decoupling i.e. to understand the Sachs-Wolfe effect in terms of the geodesic congruence expansion in the perturbed FRW background. Moreover, Fig. \ref{fig1} suggests a relation between the shear/rotation and the linear/circular polarizations. Clearly, geodesics are purely determined by the spacetime geometry and photon propagation in a gravitational field is polarization independent (in the geometrical optics approximation). Nevertheless, as we will see, the geometric description of the CMB photon congruences turns out to be very similar to the description of the polarization.
 
To elaborate on this briefly, let us recall that the polarization state of an electromagnetic radiation (without circular polarization) in flat space can be characterized by
\be\label{int}
I_{ij}=\left<E_i E_j\right>,
\ee
which lives in the two dimensional plane perpendicular to the direction of propagation. Here, one averages over a sufficiently large time interval compared to the radiation frequency. For any given unit vector $\hat{e}^i$, $I(\hat{e})=\hat{e}^i\hat{e}^j I_{ij}$ gives the energy density stored in the $\hat{e}^i$ component of the electric field. For completely unpolarized light one has $I_{ij}=\fr12 I_0 \d_{ij}$, where $I_0$ is the total intensity (note that an unpolarized light is described by a density matrix rather than by a pure state). Therefore one may introduce a dimensionless polarization tensor 
\be\label{pol}
P_{ij}=\fr{1}{I_0}\left(I_{ij}-\fr{1}{2}\,I_0\, \d_{ij}\right),
\ee
which identifies the polarization state of the radiation. The two independent entries of $P_{ij}$ are called Stokes parameters, which are usually denoted by $Q$ and $U$. They measure the intensity difference between two orthogonal polarizations like $(\hat{x},\hat{y})$ and $(\hat{x}+\hat{y},\hat{x}-\hat{y})$, respectively.  

We will see that a similar geometric description arises for the cross sectional area of a null geodesic congruence. Namely, the metric induced on the two dimensional transverse plane plays the role of \eq{int}. As the surface deforms under shear and rotation, a nontrivial intensity profile arises, which is similar to a polarization pattern. The expansion, shear and rotation of the null geodesic congruence quantify how the metric, and thus the intensity profile, changes. To illustrate this with an example, assume that initially the photons are uniformly distributed on the circle in Fig \ref{fig1}. This corresponds to a uniform intensity profile which can be compared to an unpolarized light. When the circle is deformed to an ellipse by the shear, the photons are squeezed in the narrower side and they are diluted in the other perpendicular direction yielding a nontrivial intensity profile, which is reminiscent of a light polarized along the minor axis of the ellipse. Note that the surface brightness, which depends on the total number of photons per unit area, will still be the same if the area of the circle does not change when it turns into the ellipse. 

Clearly such an effect is caused by gravitational lensing, which has been studied extensively for the CMB photons, see e.g. \cite{glcmb} for a review and \cite{ob1,ob2} for observational results. In these previous studies, one mainly determines the deflection of the CMB photons by the scalar metric perturbations after decoupling, which becomes a total gradient. The CMB temperature map must be reconstructed by taking into account the difference between the lensed and unlensed directions. Likewise, one can also work out the shearing of the CMB photons which is potentially observable. In this approach, the derivative of the deflection angle on the sphere defines the so called magnification matrix \cite{gd3}, whose components can be identified as the expansion and shear (the rotation vanishes since the deflection is pure gradient). One can further define the CMB ellipticity from the curvature matrix of the second order derivatives of the temperature field on the sphere whose distribution can be determined from the Gaussian statistics of the temperature fluctuations \cite{el1}. The shear is known to modify this distribution in a nontrivial way \cite{el2,el3}. Since the temperature fluctuations are already first order in cosmological perturbations, these lensing effects appear at second order. 

The lensing caused by the primordial gravitational waves can also be studied similarly \cite{gw1,gw2,gw3}. In that case the deflection is not pure gradient but still the shear can be decomposed into a  gradient and a transverse part on the sky \cite{stb} where the gravitational waves only contribute to the later \cite{gw1}, but this effect turns out to be too small to be observed. 

In this paper, we reconsider the weak gravitational lensing of the CMB photons by using the geodesic deviation equation in the presence of the scalar and tensor perturbations. Unlike previous studies, which mainly calculate the change of the angular position of a single null geodesic due to lensing, we determine the geometry of the congruence, i.e. the induced geometry on the two dimensional transverse cross section of a photon beam, by explicitly working out the propagation of an initially fixed deviation vector. As it turns out, the induced transverse metric is fixed to the first order by the integrals of the primordial scalar and tensor perturbations along the unperturbed geodesic line. We show that the corresponding rotation vanishes despite the fact that the deflection is not a pure gradient in the presence of the tensor perturbation and we determine the expansion and the shear. Since the expansion of the congruence changes the intensity of the beam, one would anticipate it to be related to the Sachs-Wolfe effect and we will show this is indeed the case by deriving the effect from the solutions of the null geodesic equation (the derivation of the Sachs-Wolfe effect from the geodesic equation was first given in \cite{rd}; yet our approach has slight differences which we believe worth to present). We argue that the CMB temperature map must be taught to live not on the exact sphere of the line of sight but on the geometry determined by the induced transverse beam metric. We show that this nontrivial CMB temperature profile can be characterized by three variables analogous to the Stokes parameters of the polarization tensor. To first order these are given by the line integrals of the scalar and tensor perturbations along the unperturbed null geodesic trajectories from the time of decoupling. We discuss possible implications of our findings by comparing to earlier work. 

\section{Null geodesic congruences}

In this section we will review the motion of null geodesic congruences in a general curved spacetime. Although these results are well known (see e.g. \cite{ps,gd1,gd2,gd4,gd5}), we will emphasize some salient features which will help us to discuss the Sachs-Wolfe effect and more importantly to obtain the induced metric for the CMB intensity profile after decoupling. Let $p^\m$ denotes a vector field representing an affinely parametrized null geodesic congruence
\be
p^\m p_\m=0,\hs{7} p^\m\nabla_\m \,p^\n=0.
\ee
We use $v^\m$ to denote a deviation vector connecting two infinitesimally nearby geodesics, therefore obeying
\be\label{lt}
{\cal L}_p v^\m=[p,v]^\m=0,
\ee
where ${\cal L}$ is the Lie derivative and the square brackets imply a commutator. In the vicinity of a given geodesic one may introduce coordinates $x^\mu=x^\m(\l,v,..)$ so that 
\be
p^\mu\del_\m=\fr{\del}{\del \l},\hs{7}v^\mu\del_\m=\fr{\del}{\del v}.
\ee
The deviation vector $v^\m$ can be fixed initially at some point on a given geodesic and then it can be Lie transported by \eq{lt}. Note that 
\be
p^\m \nabla_\m\left(p^\n v_\n\right)=0,
\ee
i.e. $p^\n v_\n$ is constant along the geodesic flow. The Lie transport equation \eq{lt} can be rewritten as
\be\label{7}
p^\m\nabla_\m v_\n=(\nabla_\m p_\n)v^\m,
\ee 
so the tensor $\nabla_\m p_\n$ fully determines how a general deviation vector alters along the geodesic. 

It is easy to see that the Lie transport preserves linear independence since it is actually a diffeomorphism map. If so, one may introduce three independent deviation vectors which are perpendicular to the geodesic flow obeying $p^\m v_\m=0$. These three deviation vectors define a subspace with a single null and two space-like directions. In general, an initially space-like deviation vector may become null while it is Lie transported and vice versa. We will see below that in our case of interest, i.e. for null geodesics which are slightly modified by scalar and tensor perturbations on a FRW spacetime, two deviation vectors, which are initially chosen to be space-like, remain space-like under Lie dragging. In the following we will assume that one can choose two space-like deviation vectors, which are denoted by $v^\m_{I}$, where $I=1,2$. Evidently this  assumption must be taken with care in the presence of strong fields or caustics.  

To define a transverse cross sectional slice of a null geodesic congruence, one must introduce a second independent null vector $k^\m$ which we choose to obey
\be\label{kp}
k^\m k_\m=0,\hs{5} p^\m k_\m=-2.
\ee
The vector $k^\m$ is not unique and different choices correspond to different slices of the congruence. One can now define 
\be
h_{\m\n}=g_{\m\n}+\fr12\, p_\m k_\n +\fr12\, k_\m p_\n
\ee
which is transverse to both null vectors $h_{\m\n}p^\n=h_{\m\n}k^\n=0$ and obeys $h^{\m}{}_\n h^\n{}_\r=h^\m{}_\r$. Apparently $h_{\m\n}$ specifies a positive definite metric on a two dimensional subspace, which can be associated with the tangent space of the $k^\m$ slicing. 

The basic geometrical evolution of a specified transverse slice is fixed by the following tensor
\be\label{b} 
B_{\m\n}= h_\m{}^\r\, h_\n{}^\s \,\nabla_\s p_\r.
\ee
Recall that $\nabla_\s p_\r$ determines how a deviation vector changes along the geodesic by \eq{7} and in \eq{b} one simply projects this tensor onto the transverse slice. One may decompose $B_{\m\n}$ as 
\be\label{bdec}
B_{\m\n}=\fr12 \th h_{\m\n}+\s_{\m\n}+\o_{\m\n},
\ee
where $\o_{\m\n}=B_{[\m\n]}$, $\s_{\m\n}=\s_{(\m\n)}$, $h^{\m\n}\s_{\m\n}=0$ and $\th=h^{\m\n}B_{\m\n}$ (note that $h^{\m\n}h_{\m\n}=2$). The variables $\th$, $\s_{\m\n}$ and $\o_{\m\n}$ are called the expansion, the shear and the rotation of the congruence, which are basically pictured in Fig. \ref{fig1}. 

One should note that while the shear and rotation depend on the choice of $k^\m$, the expansion can be written as $\th=\nabla_\m p^\m$ and it is $k^\m$ independent (one can prove similarly that $\s_{\m\n}\s^{\m\n}$ and  $\o_{\m\n}\o^{\m\n}$ also do not depend on the choice of $k^\m$). Moreover, it is possible to show that if $\o_{\m\n}=0$ for one choice of $k^\m$ obeying \eq{kp} then it vanishes for any other choice, see e.g. \cite{ps}. 

To have a better understanding of these variables, let us take two space-like deviation vectors $v_I^\m$, $I=1,2$. As pointed out above, these can be chosen to obey $v_I^\m p_\m=0$, yet they are not necessarily perpendicular to $k^\m$ (even if they are initially chosen to be so). To study the geometry of the slice which is both normal to $p^\m$ and $k^\m$ one may project $v_I^\m$ to define
\be\label{pr}
\tilde{v}_I^\m=h^{\m}{}_\n v_I^\n.
\ee
Obviously, $\tilde{v}_I^\m$ belongs to the two dimensional subspace which is normal to both $p^\m$ and $k^\m$ and it can be imagined to form an (infinitesimal) parallelogram. This parallelogram can be chosen freely at some initial point, by picking out $v_I^\m$, and its subsequent evolution is completely determined by Lie dragging of $v_I^\m$ and the projection \eq{pr}. Introduce now the following two dimensional  metric 
\be\label{hij}
h_{IJ}= h_{\m\n}\,\tilde{v}_{I}^\m\, \tilde{v}_{J}^\n,
\ee
which is positive definite (because $\tilde{v}_1^\m$ and  $\tilde{v}_2^\m$ are linearly independent and space-like). This metric defines the geometry of the  parallelogram formed by $\tilde{v}_I^\m$, for instance its area is given by $A=\sqrt{\det h_{IJ}}$. One may note that 
\be\label{hij2}
h_{IJ}=h_{\m\n}v_{I}^\m v_{J}^\n=g_{\m\n}v_{I}^\m v_{J}^\n.
\ee
To determine how the geometry changes along the geodesic flow one may take its directional derivate and introduce the following decomposition
\be\label{defsh}
p^\m\nabla_\m h_{IJ}= \th\, h_{IJ}+2 \,\s_{IJ}
\ee
where $h^{IJ}\s_{IJ}=0$ and $h^{IJ}$ is the inverse of $h_{IJ}$. It is easy to see that the $\th$ defined in this equation is the same with the one introduced in \eq{bdec} and 
\be
\s_{IJ}=\tilde{v}_I^\m\,\tilde{v}_J^\n\,\s_{\m\n}=v_I^\m\,v_J^\n\,\s_{\m\n}.
\ee
Moreover
\be\label{tha}
\th= \,p^\m\nabla_\m\ln\left(\sqrt{\det h_{IJ}}\right)=\fr{d\ln A}{d\l},
\ee
i.e. $\th$ gives the rate of change of $\ln(A)$ along the geodesic flow. Similarly, one may see that $\s_{IJ}$ describes the deformations in the transverse geometry that do not alter the area of the slice (one may work out an analogous construction for the rotation parameter $\o_{\m\n}$). 

It is possible to study the geometry of geodesic congruences by defining a Sachs basis and a Jacobi map. For that one can introduce two orthonormal basis vectors $E^\m_I$ which are parallelly transported along the geodesics and are also chosen to be normal to the geodesic tangent so that
\be
g_{\m\n}E^\m_IE^\n_J=\d_{IJ},\hs{5}p^\m\nabla_\m E^\n_I=0,\hs{5}E^\m_I p_\m=0.
\ee 
Given the Sachs basis $E^\m_I$ and $p^\m$, one can construct a unique null vector $k^\m$ normalized so that $p^\m k_\m=-2$. By construction $k^\m$ must also be covariantly constant along the geodesic flow $p^\m\nabla_\m k^\n=0$, which defines a specific slicing of the congruence.

One can study the evolution of the deviation vector $v^\m_I$ along the congruence by referring to $E^\m_I$. This can be achieved by introducing the Jacobi matrix as
\be\label{j1}
v^\m_I=D_I^J E^\m_J+Y_I\, p^\m.
\ee
Note that the $Y_I$ coefficient must also be introduced since $v^\m_I$ is not necessarily orthogonal to $k^\m$. Using \eq{j1} in \eq{hij} gives the transverse beam metric in terms of the Jacobi matrix 
\be\label{jhd}
h_{IJ}=D_I^K D_J^L \d_{KL}. 
\ee
The change of the Jacobi matrix along the geodesic can be parametrized by the deformation matrix $S$ as 
\be\label{jhd2}
p^\m\nabla_\m D_I^J=\fr{d}{d\l}D_I^J= S_I^K D_K^J.
\ee
Equations \eq{jhd} and \eq{jhd2} can be compactly written using the matrix notation as $h=DD^T$ and $\dot{D}=SD$. 

One can decompose $S$ into trace, symmetric trace free and antisymmetric parts (see e.g. \cite{gd5})
\be\label{js}
S=\left(\begin{array}{cc}\hat{\s}_1+\hat{\th} & \hat{\s}_2+\hat{\o} \\  \hat{\s}_2-\hat{\o} & -\hat{\s}_1+\hat{\th}\end{array}\right),
\ee
which defines the expansion $\hat{\th}$, the shear $\hat{\s}_I$ and the rotation $\hat{\o}$ of the Jacobi field. Obviously, these must be related to the previous parameters given in \eq{bdec}. For example, by taking the derivative of \eq{jhd} and using \eq{defsh} one may obtain
\be\label{j23}
\dot{h}=\th h+2\s=S h+h S^T.
\ee
Multiplying by $h^{-1}$ and taking the trace yield
\be
\th=2\hat{\th}.
\ee
Similarly, from \eq{js} and \eq{j23} one can determine $\s_I$  in terms of $\hat{\s}_I$ and $\hat{\o}$. Here, one sees that $\s_I$ mixes in general with $\hat{\o}$ which is not surprising since these are basis dependent quantities and while \eq{js} directly refers to a Sachs basis the former expression \eq{bdec} and its contractions with 
$v^\m_I$ do not. 

\section{Null geodesics on perturbed FRW backgrounds and the geometry of CMB photon beams}

Consider the following metric
\be\label{met}
ds^2=a(\h)^2\left[-(1+2\Psi)d\h^2+[(1+2\Phi)\d_{ij}+\cc_{ij}]dx^i dx^j\right],
\ee
which describes the geometry of our spacetime to an excellent approximation after decoupling (and indeed much earlier). Although the difference between the two scalar perturbations is small, i.e. $\Phi\simeq-\Psi$, we will use \eq{met} to keep our discussion general. The gauge freedom related to coordinate transformations can be utilized to enforce $\cc_{ij}$ to be transverse and trace-free but we are not going to impose these conditions either. 

One usually considers the phase space distribution function as the main physical observable and studies its evolution along the null geodesic flow, which is enough to derive the Sachs-Wolfe effect (see below). Here, instead, we explicitly work out the null geodesic equations on \eq{met} to the first order in cosmological perturbations (see \cite{ellis} which studies the geodesic deviation equation in the unperturbed FRW background). The geodesic equations can be written as
\bea
&&p^\m=\fr{d x^\m}{d\l},\nn\\
&&\fr{d p_\m}{d\l}=-\fr12 \del_\m(g^{\n\r})p_\n p_\r,\label{geo}
\eea
where the acceleration equation is expressed in terms of the geodesic cotangent $p_\m$ with its index down. The cotangent vector can be expanded as 
\bea
&&p_0=-1+\dpo,\nn \\
&&p_i=l^i+\dpi,\hs{5}\d_{ij}l^i l^j=l^il^i=1,\label{g} 
\eea
where $l^i$ is a constant unit vector and the  index $0$ refers to the conformal time coordinate $\h$ (in this paper we never lower the index on $l^i$). The zeroth order solution is fixed so that $p^\m$ is future pointing. The unit vector $l^i$ gives the spatial direction of propagation in the unperturbed FRW spacetime. The geodesic \eq{g} can be thought to carry a unit energy. To consider a more general case, one can introduce an overall multiplicative constant so that
\bea
&&p_0\to E\,p_0,\nn\\
&&p_i\to E\, p_i,\label{e}
\eea
where $E$ can be associated with the (observer dependent) photon energy. Note that these photons with different energies move on the same path. Our final results below do not depend on the value of the constant $E$, hence we set it to 1. 

One may work out $p^\m p_\m=0$ equation to solve  $\dpo$ in terms of other variables as
\be \label{g2}
\dpo=-l^i \dpi+\Phi-\Psi+\fr12 \cc_{ij}l^i l^j.
\ee
Then, the geodesic equation $p^\m\nabla_\m p_\n=0$ for $\n=i$ implies  
\be\label{epi}
\left(\fr{\del}{\del \h}+l^j\del_j\right)\dpi=\del_i\Phi-\del_i \Psi+\fr12 (\del_i\cc_{jk})l^j l^k,
\ee
where $\del_i=\del/\del x^i$. One can solve\footnote{Note that one keeps $x^i$ fixed in carrying out $\h'$ integral in \eq{pi}. Thus, when the derivative $l^j\del_j$ acts on $\d p_i$, it can be taken inside the integral. On the other hand when the derivative $\del/\del\h$ is applied, it will both act inside and produce an extra term since the upper limit of the integral is $\h$. One can see that the derivatives acting inside the integral cancel each other when $x_L$ is chosen as in \eq{xl} and the extra term coming from the  $\del/\del\h$ acting on the upper limit of the integral produces the right hand side of \eq{epi}. Since $x^i$ is kept fixed in the $\h'$ integral and the limits of integration are $x^i$ independent, the $\del_i$ derivative in \eq{pi} can freely be taken out of the integral as in the definition of the lensing potential, see e.g. \cite{glcmb}.} \eq{epi} as 
\be\label{pi}
\d p_i(x,\h)=\del_i\int_{\h_r}^{\h} d\h'\left[\Phi- \Psi+\fr12 \cc_{jk}l^j l^k\right]\left(x_L,\h'\right)+F_i(x^j-l^j\h),
\ee
where $\h_r$ is an initial time, the argument of the variables in the square bracket is $(x_L,\h')$ and $x_L$ is the spatial position on the unperturbed geodesic path at time $\h'$, i.e. 
\be\label{xl}
x_L^i=x^i-l^i(\h-\h').
\ee
The arbitrary function $F_i(x)$ in \eq{pi} gives the homogeneous solution to \eq{epi} since  
\be
\left(\fr{\del}{\del \h}+l^j\del_j\right)F_i(x^j-l^j\h)=0
\ee
and it must be fixed by the initial conditions. Let us note that along an unperturbed null geodesic whose tangent is given by $p^0=1/a^2$, $p^i=l^i/a^2$, one has
\be
\fr{d}{d\a}=\fr{1}{a^2} \left(\fr{\del}{\del \h}+l^j\del_j\right),
\ee
where $\a$ is the corresponding affine parameter. The most general perturbed geodesic solution is therefore given by \eq{g}, \eq{g2} and \eq{pi}. 

We are only interested in the photon propagation after decoupling thus we choose $\h_r$ to be the (conformal) time marking the end of recombination. We also would like to determine how an unperturbed geodesic would be modified by cosmological perturbations after  decoupling, therefore as an initial condition we set 
\be
F_i=0,
\ee
which implies
\be
\dpi(x,\h_r)=0.
\ee
These conditions ensure that one gets null geodesics in the unperturbed FRW background when the cosmological perturbations are turned off. The photons are assumed to be in thermal equilibrium at the time of decoupling, hence the constant parameters $(E,l^i)$, which can be associated with the (unperturbed) initial photon 4-momentum at the time $\h_r$, can be taken to obey a Planckian spectrum. 

It is now straightforward to calculate the expansion, shear and rotation of the congruence once a second null vector $k^\m$ obeying \eq{kp} is specified. One may expand $k^\m$ as 
\bea
k^0=1+\d k^0,\nn\\
k^i=-l^i+\d k^i
\eea
and it is easy to see that \eq{kp} only fixes $\d k^0$ and $l^i \d k^i$. As a convenient and consistent choice, which simplifies the final expressions, we set 
\be
\d k^i = \dpi. 
\ee
Then, \eq{kp} can be seen to imply
\be
\d k^0=-l^i \dpi+\Phi-\Psi+\fr12 \cc_{ij}l^i l^j,
\ee
which actually corresponds to $\d k^\m=\d p_\m$. After a tedious calculation from \eq{b} and \eq{bdec}, we find that to the first order in perturbations
\bea
&&\th=\fr{2a'}{a^3}+\fr{2a'}{a^3}l^i\dpi +\fr{1}{a^2}\left(\del_i-l^j\del_j l^i\right)\d p_i\nn\\
&&\hs{3}-\fr{2a'}{a^3}(\Phi+\Psi)+\fr{2}{a^2}(\Phi'+
l^i\del_i\Phi)\nn\\
&&\hs{3}-\fr{a'}{a^3}\cc_{ij}l^i l^j-\fr{1}{2a^2}\cc_{ij}'l^i l^j+\fr{1}{2a^2}l^i\del_i\cc_{jk}l^j l^k-\fr{1}{a^2}\del_i\cc_{ij} l^j
+\fr{1}{2a^2}\cc_{ii}'+\fr{1}{2a^2}l^j\del_j\cc_{ii},\nn\\
&&\s_{00}=0,\nn\\
&&\s_{0i}=\s_{i0}=0,\label{esr}\\
&&\s_{ij}=P_{ik} P_{jl}(\del_k\d p_l)\nn\\
&&\hs{6}-\fr12 P_{ij}\left[\del_k\d p_k-l^k\del_k l^l\d p_l-\fr12 \cc_{kk}'-\fr12 \cc_{kl}' l^k l^l+\fr12 l^k\del_k \cc_{lm}l^l l^m+\fr12 l^k \del_k \cc_{ll}-\del_k \cc_{kl}l^l\right]\nn\\
&&\hs{6}+\fr12 P_{ik}P_{jl}\cc_{kl}'-\fr12 P_{im}P_{jn}\del_m \cc_{nk}l^k-\fr12 P_{im}P_{jn}\del_n\cc_{mk}l^k+\fr12  P_{im}P_{jn}l^k\del_k \cc_{mn},\nn\\
&&\o_{\m\n}=0,\nn
\eea
where the prime on a variable denotes derivative with respect to the conformal time $\h$, $\dpi$ is given in \eq{pi} and 
\be
P_{ij}=\d_{ij}-l^i l^j
\ee
projects onto the two-dimensional spatial plane perpendicular to $l^i$. An important result that follows from this computation is that the rotation of the congruence vanishes even in the presence of tensor perturbations\footnote{Curiously, in \cite{22} it has been noted that the rotation does not contribute at first order in $N$-body simulations of the weak lensing of large scale structure even in the presence of vector or tensor perturbations. This suggests that the absence of rotation at the linear order might be valid in more general situations.} (as pointed out in the previous section  $\o_{\m\n}=0$ for all possible choices of $k^\m$ if it vanishes for only one of them). This is despite the fact that the angular deviation ceases to be pure gradient when tensor modes exist. Note that the only non-vanishing zeroth order background contribution appears in the expansion parameter, which is written as the first term in $\th$ in \eq{esr}. By definition from \eq{b}, $B_{\m\n}$ must be transverse to both $p^\m$ and $k^\m$; we explicitly verify this property by using \eq{esr} in \eq{bdec} as a crosscheck of our computation. Some of the terms in \eq{esr} drop out if one imposes $\cc_{ii}=0$ and $\del_i\cc_{ij}=0$ as gauge conditions. 

Let us now determine how the intensity of a given congruence changes along the geodesic flow. Assume that $N$ number of photons are released at time $\h_r$ from the spatial region defined by $(x^i,x^i+\D x^i)$ with momentum in the range $(p_i,p_i+\D p_i)$, where both $\D x^i$ and $\D p_i$ are sufficiently small. In time as these photons move on geodesics, the dimensions of this small rectangular phase space region change but by Liouville's theorem its volume in the phase space 
\be
\D p_1(\h) \D p_2(\h) \D p_3(\h) \D x^1(\h) \D x^2(\h) \D x^3(\h)
\ee
remains constant (see the Appendix). Defining 
\be\label{pp}
p=\sqrt{g_{ij}p^ip^j}
\ee
one has
\bea
\D p_1 \D p_2 \D p_3 \D x^1 \D x^2 \D x^3&=&(\det g_{ij})\D p^1 \D p^2 \D p^3 \D x^1 \D x^2 \D x^3\nn\\
&=&p^2\,\D p\, \D \O \,\sqrt{\det g_{ij}} \D x^1 \D x^2 \D x^3,\label{invvol}
\eea
where $\O$ refers to the solid angle \footnote{To be more specific,  defining $p^i=p n^i$, where $g_{ij}n^i n^j=1$,  the volume form in the momentum space can be expressed as $\e_{ijk}dp^i \wedge dp^j\wedge dp^k = 3 p^2 dp (\e_{ijk}n^i d n^j\wedge d n^k)=6 p^2 dp\wedge d\O$, from which the infinitesimal solid angle $d\O$ can be read.} in the tangent space about $p^i$.

The energy of these photons as observed by an observer with the proper 4-velocity vector $u^\m$ (obeying $u^\m u_\m=-1$) is given by $-u^\m p_\m$ and for a comoving observer with $u^\m=(|g_{00}|)^{-1/2}\d^\m_0$ one may define the specific intensity (or surface brightness) as
\be
I(p)= \fr{-N p_0}{|g_{00}|^{1/2}(\D p) (\D \O) V_P},
\ee
where $V_P=\sqrt{\det g_{ij}} \D x^1 \D x^2 \D x^3$ is the physical volume (which also equals $A\d t$, where $A$ is the area of measuring device and $\d t$ is the time of measurement). When $g_{0i}=0$, which is the case of our interest,\footnote{In general one can introduce a coordinate system adopted to the observer so that at the point of observation $g_{\m\n}=\h_{\m\n}$. However we are interested in how the intensity changes along the congruence thus we imagine a family of observers.} one finds from \eq{p0} and \eq{invvol} that (see e.g. \cite{mtw})
\be\label{ipn}
I(p)\propto N p^3.
\ee
This equation implies the well known standard rule that the gravitational lensing conserves surface brightness.   

To proceed one has to specify the initial distribution of the  photons and we assume that this depends on the initial momentum $p(\h_r)$, thus $N=N(p(\h_r))$ (in our case $N(p(\h_r))$ has the thermal spectrum). Note that all these photons move on the same trajectory since multiplying the geodesic tangent $p^\m$ by a constant corresponds to a rescaling of the affine parameter. Using \eq{g} and \eq{pp} one can find 
\be
p=\fr{C}{a}\left[1+l^i\dpi -\Phi-\fr12 \cc_{ij}l^i l^j\right]p(\h_r),
\ee
where $\dpi$ is the solution given in \eq{pi} and $C$ is the constant that ensures $p(\h)|_{\h_r}=p(\h_r)$. Using this solution in \eq{ipn} and integrating out all energies give the total intensity $I$ (per solid angle) as 
\be\label{itotal}
I=\int_0^\infty I(p) dp\propto \fr{1}{a^4}\left[1+l^i\dpi -\Phi-\fr12 \cc_{ij}l^i l^j\right]^4.
\ee
This final result is valid for any initial distribution function $N(p(\h_r))$ as long as the momentum integral in \eq{itotal} converges. For the thermal spectrum $I\propto T^4$ and this yields 
\be\label{swi}
T\propto \fr{1}{a}\left[1+l^i\dpi -\Phi-\fr12 \cc_{ij}l^i l^j\right].
\ee
One can check, for instance by applying the derivative along the geodesic flow, that \eq{swi} is equivalent to the Sachs-Wolfe effect \eq{sw}. 

The relation between the Sachs-Wolfe effect and the solutions of the null geodesic equation was first shown in \cite{rd}, which noted that 
\be\label{jw}
\o=-u_\m p^\m= \fr{1}{a}\left[1+l^i\dpi -\Phi-\fr12 \cc_{ij}l^i l^j\right].
\ee
This has precisely the same scaling behavior with \eq{swi} and assuming that the thermal spectrum is preserved it implies the  Sachs-Wolfe effect. In our analysis we find that the Liouville's theorem implies $(\D p) (\D \O) V_P\sim 1/p^2$ and this yields the scaling of the surface brightness \eq{ipn}, which then gives the more general result \eq{itotal}  valid for any initial (and not necessarily thermal) distribution of photons.  

\begin{figure}
\centerline{\includegraphics[width=9cm]{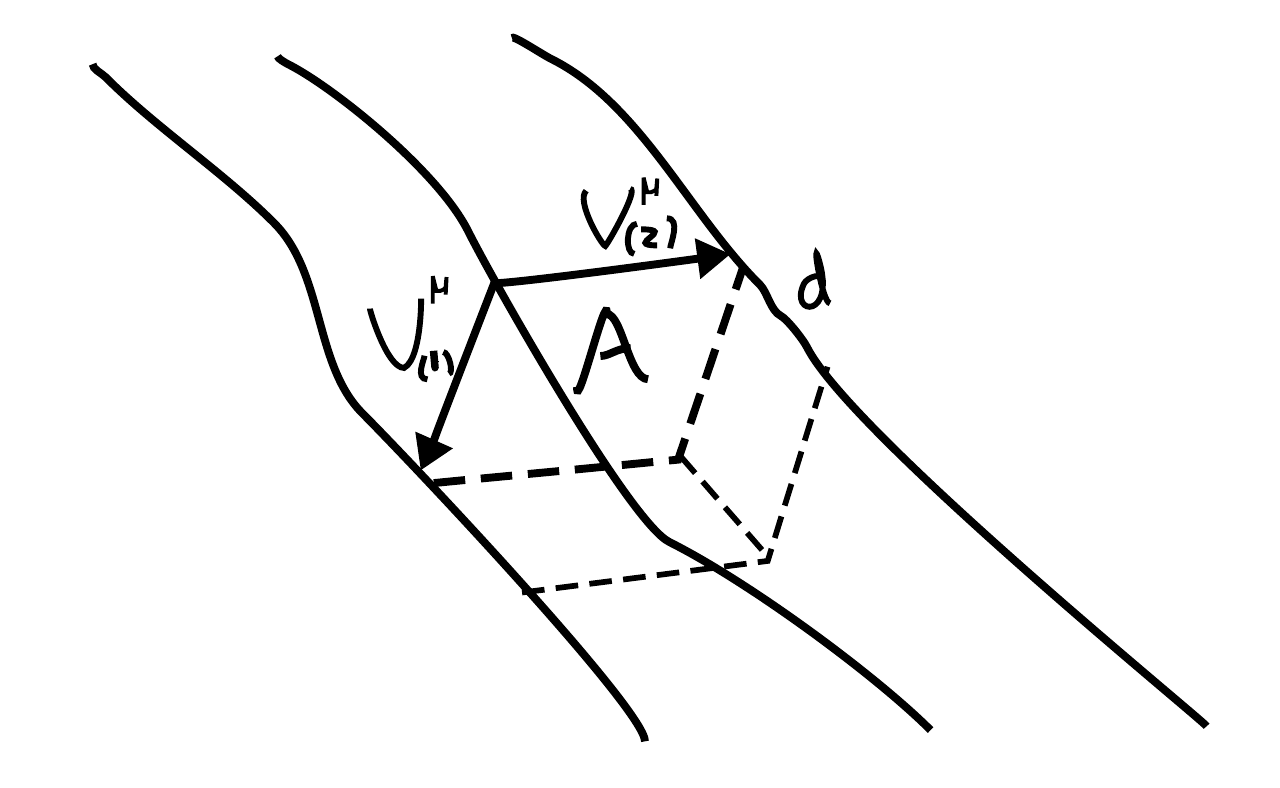}}
\caption{The sketch of null geodesic congruences with two space-like deviation vectors defining a transverse cross sectional area $A$.  One may imagine this congruence to represent the motion of a bunch of photons contained in a small physical volume $Ad$, where one extends the surface $A$ by a perpendicular length $d$.} 
\label{fig3}
\end{figure}

The above derivation does not depend on how the spatial volume containing the initial photon bunch deforms in time but it is possible to determine this evolution by considering a region that is formed by the two spatial deviation vectors and a  sufficiently small direction perpendicular to them (see Fig. \ref{fig3}). If the surface area and the normal distance are denoted by $A$ and $d$, respectively, one has 
\be  \label{51ad}
A d=\sqrt{\det g_{ij}} \D x^1 \D x^2 \D x^3.
\ee
As discussed in the previous section, see \eq{tha}, the expansion $\th$ determines how the area $A$ changes along the geodesic flow and $d$ can be related to the proper time interval $d=|g_{00}|^{1/2}\D \h$, where $\D \h$ is an initially fixed constant. 

\begin{figure}
	\centerline{\includegraphics[width=9cm]{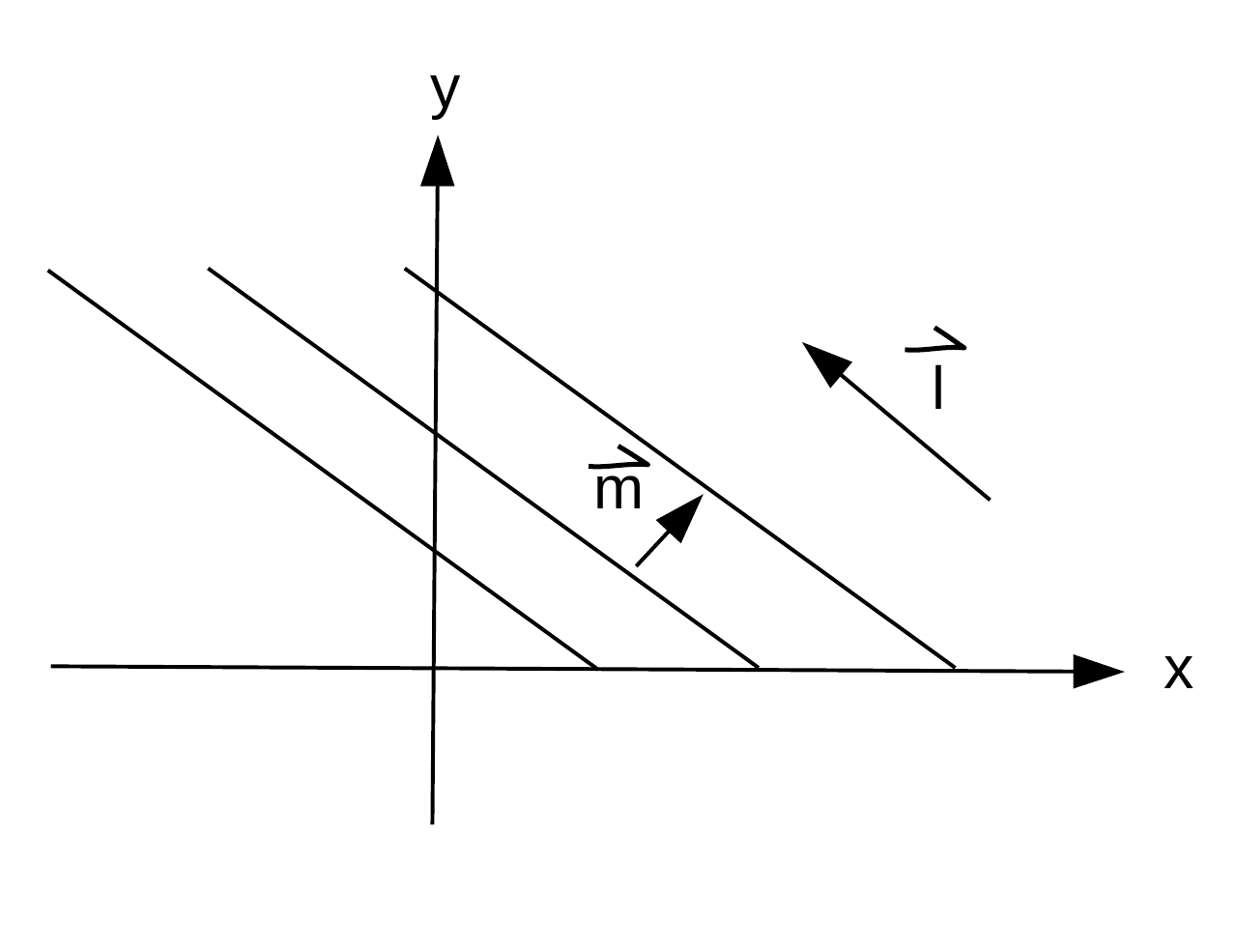}}
	\caption{The sketch of a geodesic flow parametrized by $\vec{l}$ and a deviation vector $\vec{m}$ labeling two nearby geodesics.} 
	\label{fig4}
\end{figure}

In fact, we already know more than how area changes since we have calculated the shear, which quantifies the area preserving deformations of the transverse cross section, see \eq{defsh}. Earlier studies show that the shear alters the so callled hot and cold spot CMB ellipticities \cite{el2,el3}. We can pin down the impact of the shear more directly from the induced geometry of the transverse slice. For that we explicitly work out the Lie transport of the deviation vector where the general case was reviewed in the previous section. We parametrize the deviation vector as  
\bea
&&v^0=0+\d v^0,\nn\\
&&v^i=m^i+\d v^i,
\eea
where $m^i$ is a constant normalized vector which is also perpendicular to $l^i$: 
\be\label{m}
m^i m^i=1,\hs{5} m^i l^i=0. 
\ee
To the zeroth order in perturbations, $v^\m$ is chosen to be a space-like vector and the perturbative corrections do not change this property since they are assumed to be small (which is usually granted in this context). Note that $v^\m$ connects two infinitesimally nearby geodesics (see Fig. \ref{fig4}). As in the previous section we impose $v^\m p_\m=0$, which can be used to determine $\d v^0$ as   
\be\label{v0}
\d v^0=l^i\d v^i +m^i\dpi -\cc_{ij}m^i l^j.
\ee
Then, the Lie transport equation \eq{lt} yields 
\be\label{vi}
\left(\fr{\del}{\del \h}+l^j\del_j\right)\d v^i=-\fr{2a'}{a}l^i\left( l^j \d v^j+m^j\d p_j -\cc_{jk}m^j l^k\right)+m^j\del_j\left(\dpi-2\Phi l^i -\cc_{ij}l^j\right),
\ee
where $\dpi$ is given in \eq{pi}. 

One can define two linearly independent deviation vectors satisfying \eq{m} and thus we introduce another constant vector $n^i$ obeying  
\be\label{n}
n^i n^i=1,\hs{5} n^i l^i=0, \hs{5} n^i m^i=0
\ee
so that $(l^i,m^i,n^i)$ form an orthonormal basis (in the Euclidean space with the metric $\d_{ij}$). We will respectively label the corresponding vectors as 
\be
v_1^\m=v_1^m(m^i),\hs{5}v_2^\m=v_2^\m(n^i).
\ee
The vector $v_2^\m$ obeys the same equations \eq{v0} and \eq{vi}, where one replaces $m^i$ with $n^i$. These two deviation vectors determine a transverse two dimensional cross section of the null geodesic congruence. It is now straightforward to calculate the components of the metric \eq{hij}, by utilizing \eq{hij2}, as  
\bea
&&h_{11}=v_1^\m v_1^\n g_{\m\n}=a^2\left(1+2\Phi+\cc_{ij}m^i m^j+2m^i\d v^i_1,\nn\right)\\
&&h_{22}=v_2^\m v_2^\n g_{\m\n}=a^2\left(1+2\Phi+\cc_{ij}n^i n^j+2n^i\d v^i_2,\right),\label{h12}\\
&&h_{12}=h_{21}=v_1^\m v_2^\n g_{\m\n}=a^2\left(\cc_{ij}m^i n^j+m^i\d v_2^i +n^i \d v_1^i\right).\nn
\eea
Note that the metric takes the form
\be\label{dhij}
h_{IJ}=a^2(\d_{IJ}+\d h_{IJ}).
\ee
and the off-diagonal component does not acquire a zeroth order background contribution. After solving \eq{vi} for $\d v^i_{1}$ and 
$\d v^i_{2}$ one can determine $\d h_{IJ}$ as
\bea
\d h_{11}(x,\h)&=&2\Phi(x,\h)+\cc_{ij}(x,\h)m^i m^j-2m^i\del_i \int_{\h_r}^\h d\h'\cc_{jk}(x_L,\h')m^j l^k\nn\\
&+&2(m^i\del_i)^2\int_{\h_r}^\h d\h'(\h-\h')\left[\Phi-\Psi+\fr12\cc_{jk}l^j l^k\right](x_L,\h')\nn\\
\d h_{22}(x,\h)&=&2\Phi(x,\h)+\cc_{ij}(x,\h)n^i n^j-2n^i\del_i \int_{\h_r}^\h d\h'\cc_{jk}(x_L,\h')n^j l^k\nn\\
&+&2(n^i\del_i)^2\int_{\h_r}^\h d\h'(\h-\h')\left[\Phi-\Psi+\fr12\cc_{jk}l^j l^k\right](x_L,\h')\label{dh12}\\
\d h_{12}(x,\h)&=&\cc_{ij}(x,\h)m^i n^j-m^i\del_i \int_{\h_r}^\h d\h'\cc_{jk}(x_L,\h')n^j l^k-n^i\del_i \int_{\h_r}^\h d\h'\cc_{jk}(x_L,\h')m^j l^k\nn\\
&+&2n^i m^j \del_i \del_j \int_{\h_r}^\h d\h'(\h-\h')\left[\Phi-\Psi+\fr12\cc_{kl}l^k l^l\right](x_L,\h'),\nn
\eea
Among the terms that do not appear inside the integrals, $2\Phi(x,\h)$ gives the monopole contribution, which depends only on the invariable observation point and thus cannot be extracted. The others $\cc_{ij}(x,\h)m^i m^j$, $\cc_{ij}(x,\h)n^i n^j$ and $\cc_{ij}(x,\h)m^i n^j$ vary with the direction of observation but their contribution must be negligibly small as they depend on the current amplitude of the primordial gravitational waves. The explicit $(\h-\h')$ factors arise from the following integral identity
\be\label{66}
\int_{\h_r}^{\h}d\h'' \int_{\h_r}^{\h''} d\h' \,G(x^i-l^i(\h-\h'),\h')= \int_{\h_r}^{\h} d\h'(\h-\h')\,G(x^i-l^i(\h-\h'),\h'),
\ee
which can easily be proved by changing the orders of $\h'$ and $\h''$ integrations. One can see that 
\be
\ln \sqrt{\det h_{IJ}}\simeq 2\ln a +\fr12 \cc_{ij}m^i m^j+\fr12 \cc_{ij}n^i n^j + 2\Phi+m^i\d v_1^i + n^i\d v_2^i
\ee
and using this in \eq{tha} yields $\th$ in \eq{esr}, which is a crosscheck of our computations. 

The shear appears since there is a difference between the directions 1 and 2 in \eq{h12}. One may  define
\be\label{q}
\fr{Q}{I_0}=\sqrt{\fr{h_{11}}{h_{22}}}-1\simeq m^i\d v_1^i - n^i \d v_2^i +\fr12 \cc_{ij}m^i m^j -\fr12 \cc_{ij}n^i n^j,
\ee
which measures the difference of the lengths of $v_1^\m$ and $v_2^\m$ (here $I_0$ denotes the average intensity of the beam). To read the size mismatch between two other independent directions, which is enough to completely fix the deformation at the linearized level, we introduce 
\be 
u_{1,2}^\m=v_1^\m\pm v_2^\m
\ee
and 
\be\label{imet}
\tilde{h}_{IJ}=u_I^\m u_J^\n g_{\m\n}. 
\ee
We then define 
\be\label{uu}
\fr{U}{I_0}=\sqrt{\fr{\tilde{h}_{11}}{\tilde{h}_{22}}}-1\simeq m^i\d v_2^i + n^i \d v_1^i +\cc_{ij}m^i n^j, 
\ee
which measures the length difference between $u_{1}^\m$ and $u_{2}^\m$. Neglecting the monopole and other small contributions as pointed out below \eq{dh12}, we find 
\bea
&&\fr{Q}{I_0}=-m^i\del_i \int_{\h_r}^\h d\h'\cc_{jk}(x_L,\h')m^j l^k+n^i\del_i \int_{\h_r}^\h d\h'\cc_{jk}(x_L,\h')n^j l^k\nn\\
&&\hs{6}+\left[(m^i\del_i)^2-(n^i\del_i)^2\right] \int_{\h_r}^\h d\h'(\h-\h')\left[\Phi-\Psi+\fr12\cc_{jk}l^j l^k\right](x_L,\h')\nn\\
&&\fr{U}{I_0}=-m^i\del_i \int_{\h_r}^\h d\h'\cc_{jk}(x_L,\h')n^j l^k-n^i\del_i \int_{\h_r}^\h d\h'\cc_{jk}(x_L,\h')m^j l^k\nn\\
&&\hs{6}+2n^i m^j \del_i \del_j \int_{\h_r}^\h d\h'(\h-\h')\left[\Phi-\Psi+\fr12\cc_{kl}l^k l^l\right](x_L,\h'),\label{fqu}
\eea
where $x_L$ is given in \eq{xl}. The $Q$ and $U$ parameters determine the shape of the beam ($Q=U=0$ gives a circle) and they account for the infinitesimal intensity profile. Indeed it is possible to define the following tensor on the transverse cross section\footnote{Note that $Q>0$ means that the length along the direction 1 is larger than that of the direction 2, which implies that the photons are diluted more. Thus, the intensity gets smaller along the direction 1 as compared to the direction 2.}
\be\label{iij}
I_{IJ}=\left[\begin{array}{cc}I_0/2-Q & U\\ U & I_0/2+Q\end{array}\right]
\ee
which is reminiscent of the polarization tensor \eq{pol}. 

\section{Comparison to earlier work and the CMB intensity profile}

In this section we will compare our findings to earlier work and discuss their implications. The lensing of the CMB photons is usually considered by defining the deflection angle. The geodesic equation 
\be\label{ge}
\fr{dx^i}{d\l}=\fr{l^i}{a^2}+\d p^i
\ee
can be integrated out to determine the difference between the lensed and unlensed directions. Here, $l^i$ defines the (unperturbed) line of sight and the two orthonormal vectors $m^i$ and $n^i$, which are introduced in \eq{m} and \eq{n}, are tangent to the sphere defined by $l^i$. We note that the tangent and cotangent perturbations are related to each other as
\be\label{dp2}
\d p^i=\fr{1}{a^2}\dpi -\fr{1}{a^2}\left(2\Phi l^i+\cc_{ij}l^j\right)
\ee
and from \eq{pi} one sees that $\dpi$ is pure gradient (for $F_i=0$). The deflection angle $\D\th^i$ can be defined as 
\be\label{Dth}
\D \th^i=m^i (m^j\D x^j)+n^i (n^j \D x^j),
\ee
where $\D x^i$ is obtained by integrating \eq{ge} from decoupling to present time. It is easy to see from \eq{ge} and \eq{dp2} that $\D \th^i$ is not pure gradient for $\cc_{ij}\not=0$. In that case it is not possible to introduce a lensing potential and one must directly work with the deviation angle \eq{Dth}. 

The basic lensing effect is obtained by noting that it yields a modified temperature field on the sky $\tilde{T}$ by 
\be\label{ltm}
\tilde{T}(l^i)=T(l^i+\D \th^i),
\ee
where $T(l^i)$ is the unlensed map. For weak lensing, one may expand $T(l^i+\D \th^i)$ to first order in $\D\th^i$ and the statistical properties of the observable $\tilde{T}$ can be determined from that of $T$ and $\D\th^i$, where \eq{Dth} can be used to fix the power spectrum of $\D\th^i$ in terms of the metric perturbations. This effect is second order in perturbations, since the temperature fluctuations are already first order. Moreover it vanishes for an isotropic last scattering surface since it simply remaps the points on the sphere. 

To single out the gravitational wave contribution, one may define the rotation\footnote{One should not confuse $\o$ in \eq{sdw} with the rotation $\o_{\m\n}$ defined earlier. While the former is nonzero the latter vanishes as we have calculated.} of the deflection angle as 
\be\label{sdw} 
\o=\e^{ij}D_{i} (\D \th^{j})
\ee
where $D_i=m^i (m^j\del_j)+n^i (n^j\del_j)$ is the derivative operator perpendicular to $l^i$ and $\e^{ij}=m^in^j-n^i m^j$ is the two dimensional epsilon tensor. It is easy to see that with the action of the anti-symmetric derivative the scalar perturbations  drop out in $\o$. One may then introduce the power spectrum of $\o$ and the corresponding moments $C_l^{(\o)}$, see \cite{gw1}. This can in principle be detected by observing ellipticities of distant galaxies, yet the signal to noise ratio turns out to be extremely small \cite{gw1}.

\begin{figure}
\centerline{\includegraphics[width=9cm]{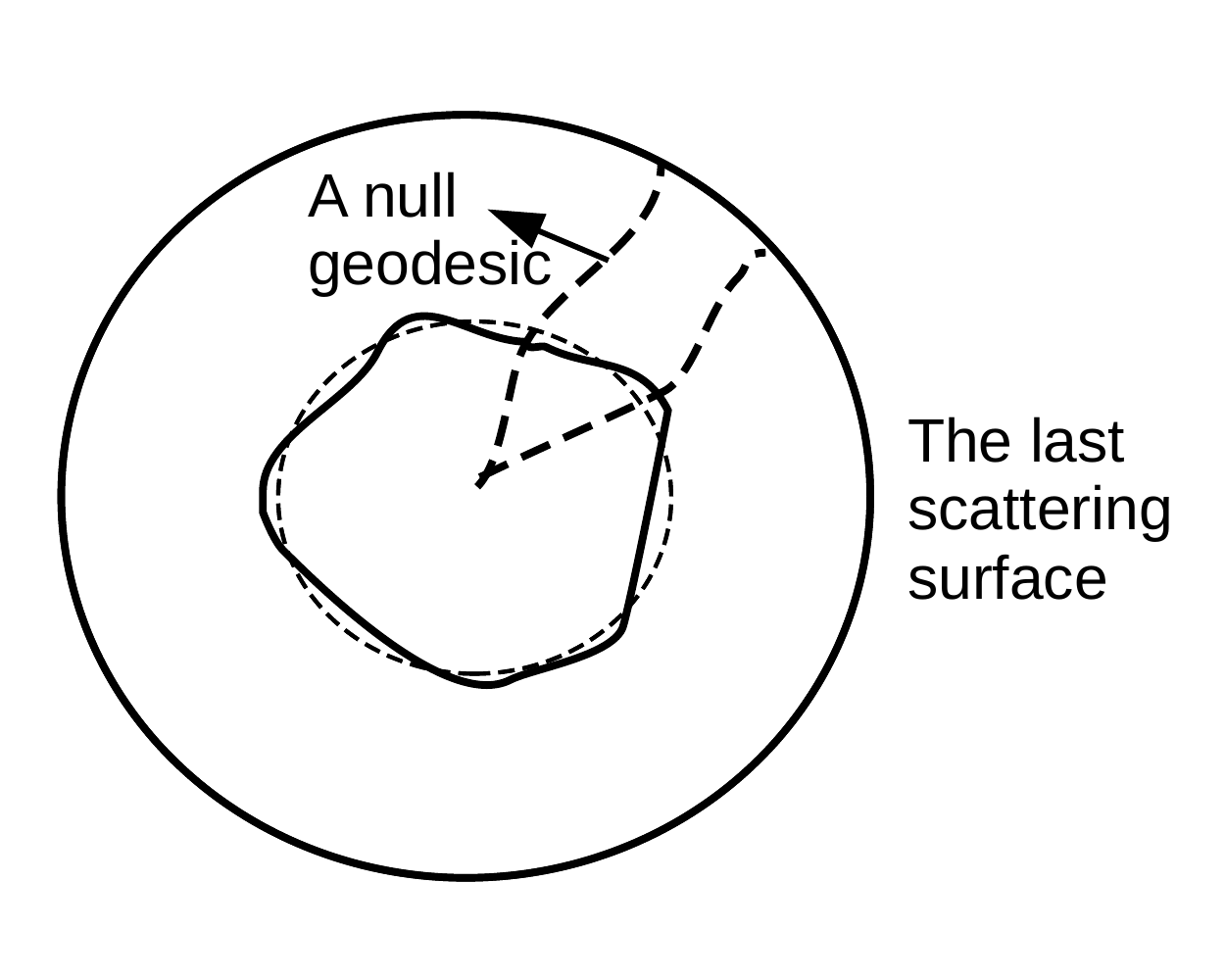}}
\caption{The sketch of the two lensed photon paths that travel from the last scattering surface. The lensing effect described by \eq{ltm} determines the photons real angular positions on the sky and the metric \eq{imet} gives the sky's nontrivial geometry which is slightly different than the round sphere indicated by the dotted lines.} 
\label{fig5}
\end{figure}

To sum up, the previous work on the gravitational lensing of the CMB photons mainly focuses on the deflection angle (where the shearing effects are extracted from its derivatives). Here, we see that a CMB photon beam, which is observed in the apparent direction $l^i$ and actually deflected by $\D \th^i$ due to lensing, has a nontrivial profile; the photons are not distributed uniformly on the transverse cross section but has an elliptical spread. This deformation is characterized by the metric \eq{imet} (or by the tensor \eq{iij}), which has small fluctuations around the spherically symmetric geometry. When different directions are considered together, the CMB temperature map must be thought to live in the deformed sky (see Fig. \ref{fig5}). This gives a much more direct and global description of the shear and without doubt this encompasses the effects related to  the hot and cold spot CMB ellipticity distribution mentioned above. 

Obviously, the beam geometry picks up first order corrections from cosmological perturbations and one may wonder why this effect has not been observed so far. As we discussed above (see \eq{51ad} and Fig. \ref{fig3}), the area of the beam cross section can be related to the Sachs-Wolfe effect which determines the variations in the surface brightness and hence the temperature fluctuations. In that case \eq{ltm} is the only deducible result and the effect necessarily emerges in the second order. 

Yet, the beam geometry contains more data than the cross sectional area of the beam (but that information does not appear in the temperature fluctuations as we pointed out). The deformations implied by \eq{fqu} show that the photons hitting a telescope's collecting area is not uniformly distributed over it. This should be difficult to measure directly, if not impossible. In principle, given any direction\footnote{One can specify $l^i$ by giving the usual spherical angles $(\th,\phi)$ and the deviation vectors $(m^i,n^i)$ can be chosen as the standard tangent vectors on the sphere  $(\hat{\th}, \hat{\phi})$, respectively.} $l^i$, one can imagine collecting photons not from a circular area but from an opening that has a sense of direction, like from slits extending along $m^i$, $n^i$, $m^i+n^i$ or $m^i-n^i$ directions. While the variable $Q$ gives the  intensity difference between the slits extending along $m^i$ and $n^i$ directions, the variable $U$ gives the difference between the slits extending along $m^i+n^i$ and $m^i-n^i$ directions. The magnitude of the signal is determined from the expectation values $\left<Q(l_1)Q(l_2)\right>$, $\left<Q(l_1)U(l_2)\right>$ and $\left<U(l_1)U(l_2)\right>$ where $l^i_1$ and $l_2^i$ are two different directions on the sky. Note that these expectation values do not depend on the photon frequency which is typical in lensing effects. Making these measurements may not be possible due to technical challenges (or the signal may turn out to be too small to be observed) but still having a proof of concept is important. 

A curious feature to note is that in the correction \eq{dh12} the scalar perturbations only appear in a double integral contrary to the tensor mode which appears in a single integral as well (recall that the explicit $(\h-\h')$ factors arise from the double integral by the identity \eq{66}). This is because after projecting to a direction perpendicular to $l^i$ in \eq{dp2}, the scalar perturbation disappears but the tensor mode survives (this is also the reason why the lensing potential does not exist in the presence of tensors). As a result, the scalar perturbations contribute only indirectly from the integral of $\dpi$ which already contains an integral \eq{pi} hence yielding a double integration. Can there be a hierarchy in the magnitudes between the single and the double integrals? It is difficult to answer this question without making an explicit calculation since there are many different factors contributing like the behavior of the mode functions or the extra $(\h-\h')$ term that may potentially give a peak somewhere in the middle of the integration interval.\footnote{I would like to thank the anonymous referee for pointing this out.} We hope to address this issue in a future work. 

\section{Conclusions}

In this paper we study the geometry of null geodesic congruences corresponding to the free streaming CMB photons on a general perturbed FRW background. We explicitly calculate the expansion, shear and rotation of the congruence as the basic geometrical parameters. We find out that the rotation vanishes even though it is not possible to define a lensing potential in the presence of tensor perturbations. We obtain the Sachs-Wolfe effect directly from the solutions of the null geodesic equation by working out how the specific intensity of the beam changes along the photon trajectory as observed by a comoving observer. Not surprisingly, it is possible to relate the expansion of the congruence to the Sachs-Wolfe effect.

As the main result of this paper, we explicitly work out the induced geometry of the transverse two dimensional cross section of the CMB photon beams. The variation of this metric along the congruence can be decomposed as the expansion and shear. We see that in the linearized case it is enough to introduce two parameters to describe the induced geometry compared to a flat circle. These are reminiscent of the Stokes polarization parameters and we determine them in terms of the primordial scalar and tensor perturbations. The CMB temperature fluctuations must be defined with respect to this induced geometry which becomes a round sphere when the cosmological perturbations are neglected. This offers a much more direct and global picture of the shear. It would be interesting to see if the present formulation, which is illustrated in Fig. \ref{fig5}, yields new observational features.  
 
\acknowledgments{This work is supported by IIE-SRF fellowship program. I am also grateful to the Harvard SAR Program and thank the colleagues at Harvard University and Connecticut College for their support and hospitality.} 

\appendix*

\section{The derivation of the Sachs-Wolfe effect}

In this appendix we review the derivation of the Sachs-Wolfe effect (with some salient differences compared to the standard derivations given in e.g. \cite{b1,b2}). We first recall how the null geodesic motion can be described as a (constraint) Hamiltonian system. It is easy to verify that the geodesic equations can be derived from the following action 
\be
S=\int d\l\left[p_\m\fr{dx^\m}{d\l}-\fr12 g^{\m\n}(x)\,p_\m p_\n\right],
\ee
which has the Hamiltonian 
\be
H=\fr12 g^{\m\n}(x)\,p_\m p_\n. 
\ee
Here one treats $(x^\m,p_\n)$ as canonically conjugate variables and  the evolution is parametrized by some $\l$, which actually corresponds to an affine geodesic parameter. For null geodesics one  would like to impose an additional constraint
\be\label{cons}
p^\m p_\m=g^{00}p_0^2+2g^{0i}p_0p_i+g^{ij}p_i p_j=0.
\ee
In that case the original action becomes reparametrization invariant, consequently one can solve the constraint \eq{cons} for $p_0$ as
\be\label{p0}
p_0=\fr{-g^{0i}p_i+\Delta}{g^{00}},\hs{5}\Delta=\sqrt{(g^{0i}g^{0j}-g^{00}g^{ij})p_i p_j}\, ,
\ee
and eliminate the dynamical variable $x^0$ as the time parameter (the root in \eq{p0} is chosen so that $p^0=\Delta>0$). This gives a reduced action for the remaining canonical variables $(x^i,p_i)$ 
\be
S=\int dx^0\left[p_i\fr{dx^i}{dx^0}+p_0\right],
\ee
where $-p_0$, which is fixed by \eq{p0} in terms of $(x^i,p_i)$, becomes the reduced Hamiltonian. 

\begin{figure}
	\centerline{\includegraphics[width=9cm]{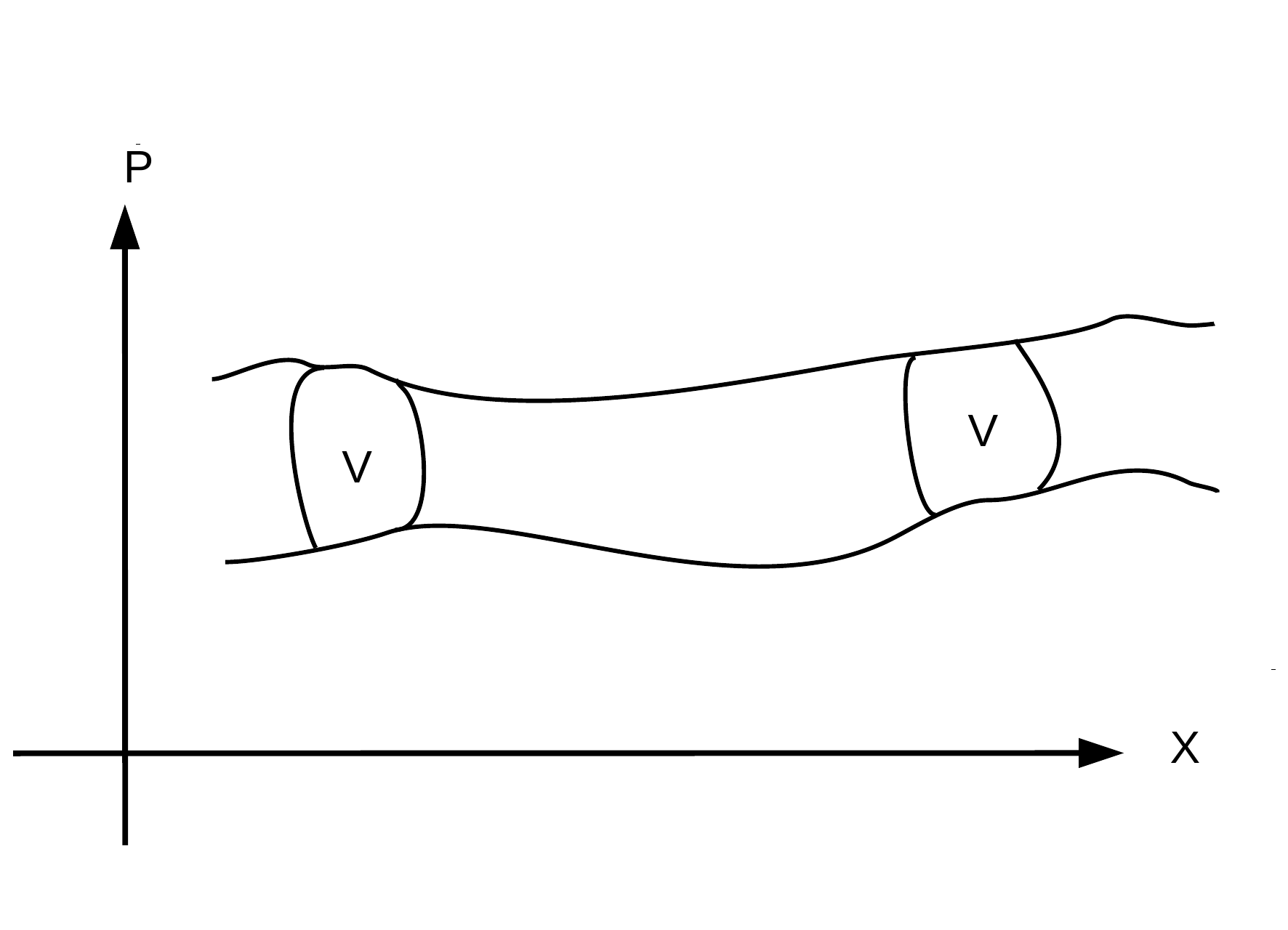}}
	\caption{The Liouville's theorem states that the volume of a region in the phase space is invariant under the Hamiltonian flow.} 
	\label{fig2}
\end{figure}

The Liouville's theorem states that the volume of a region  in the phase space $(x^i,p_j)$ does not change when it is transformed under the Hamiltonian evolution, see Fig. \ref{fig2}. Thus, an initially specified phase space distribution function $f(x^i,p_j;x^0)$ of photons must not change under the geodesic flow $df/d\l=0$. Note that the Liouville's theorem does not specify how a given region in the phase space deforms in time, it only states that its volume is constant. 

Let us now consider the photon phase space distribution function after decoupling $f(x^i,p_j;\h)$, which is assumed to have the following form
\be\label{dist}
f=\fr{2}{\exp\left[\fr{p}{T_0(1+\Theta)}\right]-1}
\ee
where $p=\sqrt{\d^{ij}p_ip_j}$, $T_0=T_0(\h)$ and the parameter $\Theta=\Theta(x^i,p_j,\h)$ describes the deviation from the Planck spectrum caused by the cosmological perturbations. When $f$ is evaluated on a null geodesic by setting $x^\m=x^\m(\l)$ and  $p^\m=p^\m(\l)$, the Liouville's theorem (in the absence of interactions) ensures that
\be
\fr{df}{d\l}=\fr{\del f}{\del x^\m}\fr{dx^\m}{d\l}+\fr{\del f}{\del p_i}\fr{dp_i}{d\l}=0. \label{fgeo}
\ee
The 4-velocity vector of a comoving observer in \eq{met} (obeying $u^\m u_\m=-1$) is given by 
\be\label{u} 
u^0=\fr{1}{a}(1-\Psi),\hs{5}u^i=0
\ee
and the energy measured by this observer is 
\be\label{w}
\o=-u^\m p_\m. 
\ee
One can use these relations to rewrite \eq{dist} in terms of the observable quantities to interpret $T_0$ and $\Th$ physically. To the zeroth order (i.e. when $\Th=0$ and one considers unperturbed geodesics in the FRW spacetime), one finds from \eq{fgeo} that $T_0$ must be a constant and from \eq{w} that $\o=p/a$. Replacing now $p$ with $\o$ in \eq{dist}, one sees that the observed temperature of the photons must be identified in the unperturbed case as
\be
T=\fr{T_0}{a}.
\ee
Similarly, to the first order in perturbations, one can find that
\be
\fr{\d T}{T}=\Th-\Phi-\fr12 \cc_{ij}l^i l^j
\ee
and \eq{fgeo} implies the Sachs-Wolfe effect
\be\label{sw}
\left(\fr{\del}{\del \h}+l^j\del_j\right)\fr{\d T}{T}=-\Phi'-l^i\del_i\Psi-\fr12 \cc_{ij}'l^il^j,
\ee
see e.g. \cite{b1,b2}.

\end{document}